\documentclass[usenatbib,useAMS]{mn2e}
\usepackage{epsfig}
\usepackage{subfigure}
\usepackage{times}
\usepackage{amsmath}
\usepackage{amssymb}
\begin{document}

\title[Star formation history and double degenerates]
{Star formation history, double degenerates and type Ia supernovae in the thin disc}

\author[S.\,Yu \& C.S.\,Jeffery]
   {Shenghua Yu\thanks{E-mail: syu@arm.ac.uk (SY)} and C. Simon Jeffery \thanks{E-mail: csj@arm.ac.uk (CSJ)}
   \\
   Armagh Observatory, College Hill, Armagh BT61 9DG,
   N. Ireland
}

     \date{Accepted .
           Received ; }

      \pagerange{\pageref{firstpage}--\pageref{lastpage}}

\label{firstpage}

    \maketitle

\begin{abstract}

We investigate the relation between the star formation history and the
evolution
of the double-degenerate (DD) population in the thin disc of the Galaxy,
which we assume to have formed 10 Gyr before the present. We introduce
the use of star-formation contribution functions as a device for
evaluating the birth rates, total number and merger rates of DDs. These
contribution functions help to demonstrate the relation between
star-formation history and the current DD population and, in particular,
show how the numbers of different types of DD are sensitive to different
epochs of star formation.

Analysis of the contribution functions given by a quasi-exponentially
decline in the  star-formation rate shows that star formation from  0 to 8
Gyr after thin-disc formation dominates the present rates and total
numbers of He+He DDs and CO+He DDs.
Similarly, the current numbers of CO+CO and ONeMg+X DDs come mainly
from early star formation ($<6$ Gyr) , although star formation from 4 to 8
Gyr continued to contribute more CO+CO than He+He DDs.
The present birth-rates for CO+CO and ONeMg+X DDs are strongly governed by
recent star formation ({\it i.e.} $8 - 9.95$ Gyr).
Star formation from $<7.5$ Gyr does {\it not} contribute to the present
birth rates of CO+CO and ONeMg+X DDs, but it has a distinct contribution
to their merger rates.

We have compared the impact of different star-formation models on the
rates and numbers of DDs and on the rates of type Ia (SNIa) and
core-collapse supernovae (ccSN).
In addition to a quasi-exponential decline model, we considered an
instantaneous (or initial starburst) model, a constant-rate model, and an
enhanced-rate model.  All were normalised to produce the present observed
star density in the local thin disc.
The evolution of the rates and numbers of both DDs and SNIa 
are different in all four models, but are most markedly different in the
instantaneous star-formation model, which produces a much higher rate than
the other three models in the past, primarily as a consequence of the
normalisation.

Predictions of the current SNIa rate range from  $\approx2$ to 
$5\times10^{-4}$ yr$^{-1}$ in the four models, and are slightly below the
observed rate because we only consider the DD merger channel. The
predicted ccSN rate ranges from 1.5 to 3 century$^{-1}$, and is consistent
with observations.

\end{abstract}


\section{Introduction}
\label{sec_introduction}

We know little about the star formation (SF) history of the Milky Way disc. Of the few ways 
available to explore the SF history, studying the stellar age distribution by the observation 
of a sample of different types of stars and comparing observed and synthetic colour-magnitude 
diagrams would be the most favored approach. The simplest view of the SF history 
would be a single star burst taking place at the formation of the disc. Substantial 
evidence, however, shows the disc has experienced a more complicated history, including repeated 
star bursts, an exponential declining SF or extended periods of enhanced SF \citep{Majewski93,Rocha-Pinto00} 
or some combination of all of these. 

The present SF rate is better understood. \citet{Smith78} 
concludes the SF rate in the Galaxy is perhaps 5 $M_{\odot}$yr$^{-1}$, 
of which 74\% take place in spiral arms, 13\% in the interarm region, and 13\% in the 
galactic center. Observations of Lyman continuum photons from O stars in giant H II regions 
give the SF rate in the disc to be 4.35 $M_{\odot}$yr$^{-1}$. A SF rate $\approx$3.6 $M_{\odot}$yr$^{-1}$ 
was suggested by \citet{McKee89} from an analysis of thermal radio emission in HII 
regions around massive stars. The measurement of the mass of $^{26}$Al in the Galaxy implies a SF rate 
$\approx$4 $M_{\odot}$yr$^{-1}$ \citep{Diehl06}. The derivation of the SF rate is strongly 
associated with the initial mass function (namely the distribution of the mass of the new-born 
stars) and the birth rate of core collapse supernovae (type Ib/c and type II). 

Double degenerates (DDs) are a type of exotic binary consisting of two 
white dwarfs. As the evolved remnants of stars, the formation of close DDs requires their progenitors to have 
undergone a mass transfer stage, either common envelope or stable Roche lobe overflow phase. 
Ultra-compact DDs are especially interesting since not only can they be 
observed as active sources of electromagnetic radiation, {\it e.g.} AM Canum Venaticorum stars, \citep{Cropper98,Ramsay05,Nelemans04}, 
but also the final products of DD evolution may include type Ia supernovae \citep{Yungelson00}). Theoretical 
results \citep{Evans87,Nelemans01b,Ruiter10,Yu10} suggest that close DDs should also be significant sources of gravitational 
wave radiation, and be detectable by the space gravitational wave (GW) detector LISA (Laser Interferometer Space Antennae). 

The census of DDs therefore is important because they allow us to test the endpoints of stellar evolution. We 
need to know the present-day numbers, birth rates, merger rates and galactic distribution 
of DDs, and also the distribution of their properties ({\it i.e.}
mass, orbital separation, chemical abundance and age). 
To interpret these, we need to investigate how the history of star formation in the Galaxy affects the 
DD population. We can achieve this using a population synthesis approach. 

In this paper, we study the correlation of the SF history and the
evolution of DD population. It may help interpret the distribution of
the physical properties of DDs as a function of the SF history. In \S
\ref{sec_population}, we describe the approach to simulate a
population of the DDs in the thin disc and discuss the details of how
a quasi-exponential declining SF rate influences the birth rate,
merger rate and total number of DDs. In \S \ref{sec_sfimpact}, we show
the distributions of some important physical parameters of the DDs
from different epochs of SF. In \S \ref{sec_comparison}, we compare
the impact of different SF models on the rates and numbers of DDs. We
discuss the results for supernovae rates in different SF models in \S
\ref{sec_snia}. Some points are discussed in \S \ref{sec_summary} and
conclusions are drawn in \S \ref{sec_conclusion}.

\section{Population synthesis of thin disc DDs in the Monte Carlo approach}
\label{sec_population}

In this section, we describe the population synthesis method 
to obtain a sample of DDs comparable with current observations, and the 
structure of the thin disc in which the DD population is distributed. 
The individual stellar-evolution tracks were computed using the method 
described by \citet{Hurley00,Hurley02}. Population synthesis was carried out 
using the method described by \citet{Yu10} in an initial study of the present 
Galactic DD population.

\begin{figure}
\centering
\includegraphics[width=12.2cm,clip,bb=27 20 580 250,angle=0]{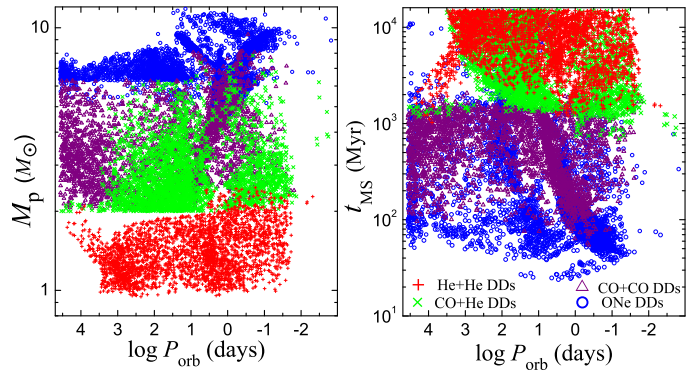}
\caption{The orbital periods 
(log $P_{\rm orb}$/day) of the DDs against the primary masses ($M_{\rm p}$, left panel) 
in the MS+MS binary progenitors and the time ($t_{\rm MS}$, right panel) 
which is from a primordial MS+MS binary to when its descendant DD was just born.}
\label{fig_progenitor}
\end{figure}

\subsection{The star-formation contribution function}
\label{sec_cfunction}

\begin{table}
%\begin{minipage}[t]{\columnwidth}
\caption{The variables in the population synthesis simulation and their relationship.}
\label{tab_variables}
\begin{center}
\begin{tabular}{lccccc}
\hline
 time \\
 $t_{\rm disc}$                                   &  the age of the thin disc  \\
 $t_{\rm sf}$                                     &  the time of star formation (SF) \\
                                                  &  ($0\leqslant t_{\rm sf}\leqslant t_{\rm disc}$) \\
 $\Delta=t_{\rm disc}-t_{\rm sf}$     &  delay time \\ 
                                                  &  from SF to the disc age \\
\hline
 Thin disc            \\
 $n(t_{\rm disc})$                          &  total number      \\
 $\dot{n}(t_{\rm disc})$                    &  number rate  \\
 $\dot{C}(\Delta)$                          &  contribution function to $n$    \\
 $\ddot{C}(\Delta)$             &  rate contribution function to $\dot{n}$     \\
 $S(t_{\rm sf})$                            &  SF rate           \\  
\hline
 Sample$^{\dag}$\\
% $n_{\ast}(t_{\rm disc})$                   &  total number in sample  \\
 $\delta n_{\ast}(\Delta)$      &  number at $t_{\rm disc}$ \\
                                            &  from SF at $t_{\rm sf}\rightarrow t_{\rm sf}+\delta t_{\rm sf}$    \\
 $\dot{C_{\ast}}(\Delta)$       &  contribution function    \\
 $\ddot{C_{\ast}}(\Delta)$      &   rate contribution function  \\
 $S_{\ast}(t_{\rm sf})$                     &  mean SF rate in the sample          \\  
 $^{\dag}$ All the values normalized by $S_{\ast}(t_{\rm sf})$.   \\  
 \hline
%\begin{tabular}{lc}
 Identities \\
 \begin{math} \dot{C}_{\ast}(\Delta)=\partial n_{\ast}(\Delta) / \partial t_{\rm sf} \end{math} \\
 \begin{math} ~~~~~~~~~~~~ \approx \delta n_{\ast}(\Delta)/ \delta t_{\rm sf} \end{math}\\
 \begin{math}
 \begin{large}
  \ddot{C}_{\ast}(\Delta)=\frac{\partial^{2} n_{\ast}(\Delta)}{\partial t_{\rm sf} \partial t_{\rm disc}} 
 \end{large}\end{math}\\
 \begin{math} \dot{C}(\Delta)=\dot{C_{\ast}}(\Delta)\cdot S(t_{\rm sf})  \end{math}\\
 \begin{math} \ddot{C}(\Delta)=\ddot{C_{\ast}}(\Delta)\cdot S(t_{\rm sf})  \end{math}\\
 \\
 $n(t_{\rm disc})=\int_{0}^{t_{\rm disc}}\dot{C}(t_{\rm disc}-t) {\rm d}t$\\
   $  ~~~~~~~~~~~~~~              =\int_{0}^{t_{\rm disc}} \dot{n}(t) {\rm d}t $\\
 \\
 \begin{math} \dot{n}(t_{\rm disc})=\int_{0}^{t_{\rm disc}}\ddot{C}(t_{\rm disc}-t) {\rm d}t  \end{math}\\
% \\
\hline
%\end{tabular}
\end{tabular}
\end{center}
%\end{minipage}
\end{table}

We define a time representing the age of the thin disc to be $t$, 
such that the disc formed at $t=0$, and the current age of the thin
disc is $t_{\rm disc}$.
During the evolution of the thin disc, we assume 
that star formation takes place over an interval $t = t_{\rm sf}$ 
($0\leqslant t_{\rm sf}\leqslant t_{\rm disc}$) to $t = t_{\rm sf}+\delta
t_{\rm sf}$.
We simulate the evolution of all binaries formed during this
interval by computing a sample (denoted by $\ast$) of $k$ 
primordial MS+MS (MS: main sequence) binaries with a total mass $M_{\ast}$\footnote{We 
make the assumption that the global properties of the MS+MS binary population from each star formation 
episode are the same; {\it i.e} each sample adopts the same initial
parameter distribution.}. 
This gives a mean {\it sample} SF rate $S_{\ast}=M_{\ast}/\delta t_{\rm sf}$. 
$\delta t_{\rm sf}$ is assumed to be small; thus all primordial MS+MS 
binaries formed during this interval evolve as though they formed at
time $t_{\rm sf}$.  The binary-star evolution tracks are computed as a 
function of $t$ over the interval $\log (t/\rm yr)= 7.4 - 10.13$ 
with $\delta \log t=0.00867$. 

We then compute the contribution of star formation during the
interval $\delta t_{\rm sf}$ to the birth-rate, 
total number and merger-rate of DDs at time $t=t_{\rm disc}$ (the present
epoch). This interval is denoted by the quantity 
$\Delta = t_{\rm disc} - t_{\rm sf}$.

For a given age of the thin disc $t_{\rm disc}$, star formation in the thin disc 
runs from $t_{\rm sf} = 0$ to $t_{\rm disc}$. For
convenience, we use $\delta \log t_{\rm sf}=\delta \log t_{\rm disc}$. 

Two additional time variables are used to establish the stage of
evolution of DDs in the sample. $t_{\rm MS}$ represents the lifetime
from birth of an individual binary system to the formation of a DD.
$t_{\rm DD}$ represents the lifetime from formation of a DD to its
merger\footnote{We have assumed the angular momentum evolution of a DD to be governed
only by gravitational wave radiation and mass transfer \citep{Yu10}, 
although tidal interaction and magnetic braking may play an important 
role \citep{Marsh04,Gokhale07,Farmer10}.}.

The birth-rate contribution is established by counting the number of DDs which
form during the interval $t_{\rm disc}$ to $t_{\rm disc}+\delta t_{\rm
  disc}$.  These are stars for which 
\begin{equation} t_{\rm disc} \leqslant t_{\rm sf}+t_{\rm MS} < t_{\rm disc}+\delta t_{\rm disc}.
\end{equation}
Normalising  by the mean sample SF rate $S_{\ast}$ gives the fractional number
  $\delta n_{\ast,\rm new}(\Delta)$ of DDs born per unit mass
of star's formed between $t = t_{\rm sf}$ and  $t_{\rm sf}+\delta
t_{\rm sf}$. 

The merger-rate contribution is established in an analagous way.
Counting stars for which 
\begin{equation} t_{\rm disc} \leqslant t_{\rm sf}+t_{\rm MS}+t_{\rm DD} < t_{\rm
   disc}+\delta t_{\rm disc},
\end{equation}
gives the fraction $\delta n_{\ast,\rm mer}(\Delta)$.

The total-number contribution $\delta n_{\ast,\rm dd}(\Delta)$
represents the fractional number of DDs in the sample which exist at
time $t_{\rm disc}$. 
These stars are indentifed by 
\begin{equation} 
t_{\rm sf}+t_{\rm MS} < t_{\rm disc} \leqslant t_{\rm sf}+t_{\rm MS}+t_{\rm DD}.
\end{equation}

These three quantities, let us call them star-formation number 
contribution functions $\delta n_{\ast}$, comprising $\delta n_{\ast,\rm new}$, $\delta n_{\ast,\rm dd}$ and
$\delta n_{\ast,\rm mer}$, may then be combined with a model for the star formation 
history of the thin disc $S(t)$ to establish the {\it total} current 
birth-rate, merger rate and number of DDs. 

From above, we have 
$\delta n_{\ast}(\Delta)$ to be the SF contribution at time 
$t_{\rm sf}$ to the number per unit SF rate at time $t_{\rm sf}$, integrated over an 
interval $\delta t_{\rm sf}$. Hence $\dot{C_{\ast}} = \partial n_{\ast}/\partial t_{\rm sf} 
\approx \delta n_{\ast}/\delta t_{\rm sf}$ 
is the sample contribution function per unit SF rate per unit time at time $t_{\rm sf}$. 

To obtain total numbers in a real thin disc, we must multiply by the thin-disc SF rate to
obtain the total contribution rates:
\begin{equation}
\dot{C}=\dot{C_{\ast}}\cdot S(t_{\rm sf})\approx 
\frac{\delta n_{\ast}}{\delta t_{\rm sf}} \cdot {S(t_{\rm sf})}.
\label{eq_cdot}
\end{equation}
$\dot{C}$ also reflects the distribution of age of the DDs. 
For example, if the thin-disc SF rate is $S(t)$, 
star formation at $t_{\rm sf}=0$ will contribute 
$\delta n_{{\ast}}(t_{\rm disc})/\delta t_{\rm sf} \cdot S(0)$ DDs with
age $t_{\rm disc}$; 
star formation at $t_{\rm sf}=1$ Gyr will contribute 
$\delta n_{{\rm ast}}(t_{\rm disc}-1)/\delta t_{\rm sf} \cdot S(1)$ DDs with
age $t_{\rm disc}-1$ Gyr, and so on. 

Since we know from Eq.\,\ref{eq_cdot} that star formation at time $t_{\rm sf}$ contributes 
$\dot{C}(t_{\rm disc}-t_{\rm sf})$ DDs at thin disc age $t_{\rm disc}$, we define a rate contribution function
\begin{equation}
\ddot{C}=\partial \dot{C}/\partial t_{\rm disc}
\label{eq_cddot}
\end{equation}
to be the contribution from star formation at $t_{\rm sf}$ to 
the number rate at time $t_{\rm disc}$. The integral of the rate contribution function 
tells us the birth rate and the merger rate of DDs. 

\subsection{Birth rate, merger rate and total number of DDs}
\label{sec_birnum}

We define $n(t)$ to be the total number of DDs at thin disc age $t$,
where $n$ may represent new-born $n_{\rm new}$, merged $n_{\rm mer}$
or existing $n_{\rm dd}$ DDs. Star formation from $t_{\rm sf}$ to $t_{\rm sf}+\delta t_{\rm sf}$ will generate 
$\dot{C}\cdot \delta t_{\rm sf}$ DDs with age $t_{\rm disc}-t_{\rm sf}$.  
Hence, the overall number of new-born, alive or 
merged DDs at $t = t_{\rm disc}$ is 
\begin{equation}
\begin{split}
n(t_{\rm disc}) 
&=\int_{0}^{t_{\rm disc}}\dot{C}(t_{\rm disc}-t_{\rm sf}){\rm d}t_{\rm sf}\\
&\approx \sum_{t_{\rm sf}=0}^{t_{\rm disc}} 
\delta n_{\ast}(t_{\rm disc}-t_{\rm sf})\cdot S(t_{\rm sf}) \\
&\approx \sum_{t_{\rm sf}=0}^{t_{\rm disc}} \dot{C}\cdot \delta t_{\rm sf}.
\end{split}
\label{eq_number1}
\end{equation}

Then we are able to calculate the number rate $\dot{n}(t_{\rm disc})$ of DDs from the contribution function $\ddot{C}$
\begin{equation}
\begin{split}
\dot{n}(t_{\rm disc})&=\int_{0}^{t_{\rm disc}} \ddot{C}(t_{\rm disc}-t_{\rm sf}) {\rm d}t_{\rm sf}\\
&\approx\sum_{t_{\rm sf}=0}^{t_{\rm disc}} \frac{\delta n_{\ast}(t_{\rm disc}-t_{\rm sf})}{\delta t_{\rm sf}}
\cdot S(t_{\rm sf}) \\
&\approx \sum_{t_{\rm sf}=0}^{t_{\rm disc}} \ddot{C}\cdot \delta t_{\rm sf},
\end{split}
\label{eq_birthrate}
\end{equation}
where $\dot{n}$ represents the birth rate $\nu$ or merger rate $\zeta$. 
We will give the details of these functions in the present simulation in \S \ref{sec_sfrDD}. 

Furthermore, since we know the overall rate of change in the number
DDs $\nu(t)-\zeta(t)$, we can also calculate the total number 
 of DDs in the thin disc at time $t_{\rm disc}$ 
($n_{\rm dd}(t_{\rm disc})$) by the integral \begin{equation}
\begin{split}
n_{\rm dd}(t_{\rm disc})=\int_{0}^{t_{\rm disc}}[\nu(t)-\zeta(t)]{\rm d}t\\
\approx \sum_{t=0}^{t_{\rm disc}} [\nu(t)-\zeta(t)] \cdot \delta t.
\end{split}
\label{eq_number2}
\end{equation}

Consequently, we have two different methods to compute the current
number of DDs in the thin disc.
Eq.\,\ref{eq_number1} represents the sum of contributions from each individual SF 
epoch to the total number by counting which
DDs exist at the current epoch. 
Eq.\,\ref{eq_number2} represents the integral of the birth-rate minus
the merger-rate, or nett birth rate, over the entire SF history of the galaxy.
The two methods should give the same result but, 
due to the limited grid in $t_{\rm sf}$, the evaluation of the sum in Eq.\,\ref{eq_number1} 
gives slightly higher numbers than Eq.\,\ref{eq_number2}. In the present simulation with 
quasi-exponential SF rate, the 
differences are 1.57\%, 0.98\%, 0.51\%, and 0.46\% for 
He+He, CO+He, CO+CO, and ONeMg+X DDs, respectively.

We list the time variables, numbers (rates), contribution functions and their relation 
in the thin disc and our simulation in Table\,\ref{tab_variables}.

\subsection{Input to the Monte-Carlo simulation}
\label{sec_ps}

The SF rate is assumed to be given by the quasi-exponential function 
\begin{equation}
S(t_{\rm sf})=7.92 e^{-(t_{\rm sf})/\tau}+0.09(t_{\rm sf})~~{\rm
    M_{\odot}yr^{\rm -1}}
\end{equation}
where $\tau=9$ Gyr \citep{Yu10}, which produces $\approx$ 3.5 $M_{\odot}$yr$^{-1}$ at the current epoch.
This is consistent with \citet{Diehl06}.

The contribution function for each starting epoch ($t_{\rm sf}$) is
obtained from a sample of primordial binaries. Each
binary is defined by at least five initial parameters: primary mass $M_{\rm p}$,
mass-ratio $q$, orbital separation $a$, eccentricity $e$ and
metallicity $Z$. Primary
masses $M_{\rm p}$ are distributed assuming a power law \citep{Kroupa93} for the initial mass function
(IMF). This is constrained by the observation 
of the local luminosity function and stellar density of \citet{Wielen83} 
and \citet{Popper80}. 
A flat  distribution is adopted for mass-ratio ($0<q<1$) and
eccentricity ($0<e<1$), since these  are not well constrained by
observation. 
The distribution of orbital separation 
$a$ is assumed to be constant in logarithm for wide binaries and falls off 
smoothly at close separations \citep{Han98}. We have assumed $Z=0.02$
throughout. All other input parameters are as given by \citet{Yu10}.

The present population synthesis simulation started with a sample of 
$10^{7}$ primordial MS$+$MS binaries, with total mass 
$\approx1.05\times10^{7}~\rm M_{\odot}$, yielding a sample total of 
$4.93\times10^{4}$ DDs over all orbital periods of less than 100
yr during a sampling period of 15 Gyr. 

In order to evaluate the computation of our DD sample, 
Fig.\,\ref{fig_progenitor} shows (left) the orbital periods 
(log $P_{\rm orb}$/day) of each new-born DD against primary mass 
($M_{\rm p}$) in the 
progenitors and (right) the time $t_{\rm MS}$ from binary-star formation 
to the birth of its descendant DD. We define DDs in 
this sample as unevolved DDs. This figure shows the relation of new-born DDs
and their progenitors. 

\subsection{The contribution functions in the present simulation}
\label{sec_sfrDD}

\begin{figure}
\centering
\includegraphics[width=15cm,clip,bb=37 10 750 280,angle=0]{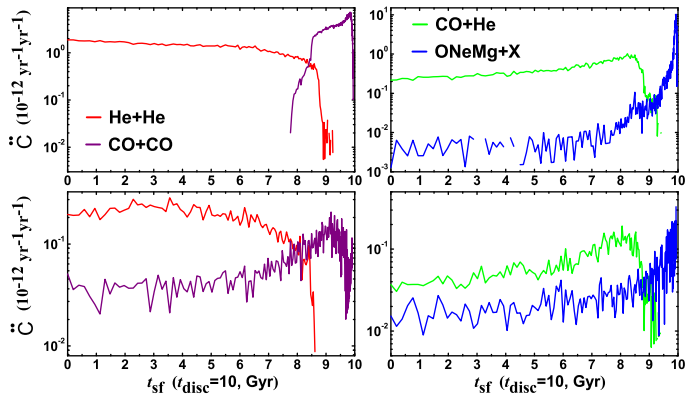}
\caption{The variation of the contribution of star formation $t_{\rm sf}$ 
to the present birth rate and merger rate of DDs, $\ddot{C}$, in the quasi-exponential 
star formation rate model. In this figure, top panels represent the contribution 
to the present birth rate, and bottom panels represent the contribution to the merger rate. 
See \S\ref{sec_sfrDD} for the details.}
\label{fig_cddot}
\end{figure}

\begin{figure}
\centering
\includegraphics[width=15cm,clip,bb=37 10 750 280,angle=0]{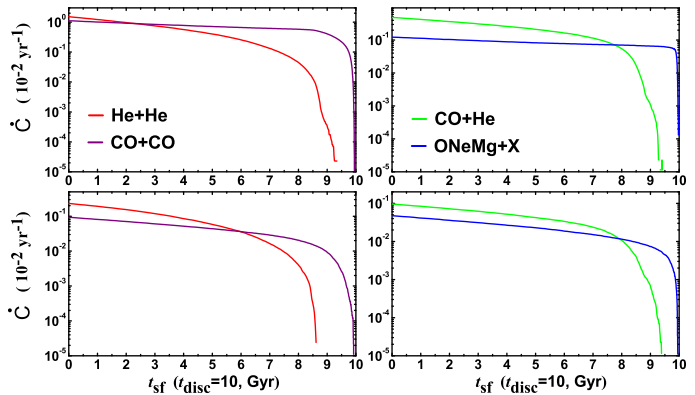}
\caption{Same as Fig.\,\ref{fig_cddot} but for the contribution of star formation 
$t_{\rm sf}$ 
to the present number (top panels) and the total merged number (bottom panels) of DDs, 
$\dot{C}$. See \S\ref{sec_sfrDD} for the details.}
\label{fig_cdot}
\end{figure}

\begin{figure}
\centering
\includegraphics[width=12.5cm,clip,bb=27 20 580 250,angle=0]{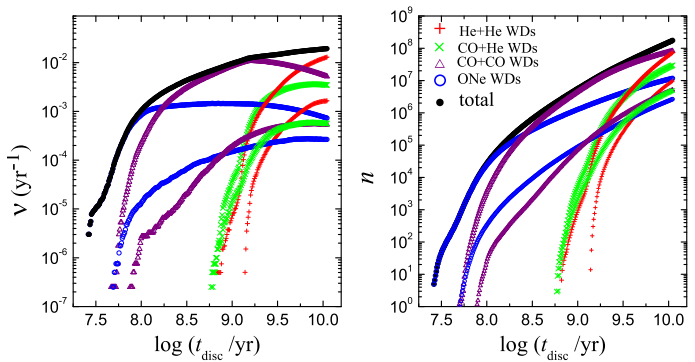}
\caption{Birth rates $\nu(t)$ (left panel) and merger rates $\zeta(t)$
  (left panel) and numbers $n(t)$ (right panel) of different types of DD by integral of 
the corresponding birth and merger rates versus the age of the thin disc $t_{\rm disc}$. 
$\nu(t)$ and $\zeta(t)$ for each type of DD is shown in the same
  colour, but $\zeta(t)<\nu(t)$. Black points are for the nett total birth rate ($\nu(t)-\zeta(t)$) 
and the total existing number of the DDs.}
\label{fig_birnum}
\end{figure}

In this section, we investigate how the present number of DDs and the number of 
new-born DDs depends on the star formation history of the thin disc by examining 
the contribution functions. We take the current age of the thin disc to be 
10 Gyr. Star formation time $t_{\rm sf}$ runs from 0 to 10 Gyr. 
Figures\,\ref{fig_cddot} and \ref{fig_cdot} illustrate the variation of $\ddot{C}$ (rate
contributions to the present birth and merger rates) and 
$\dot{C}$ (number contributions to the present numbers and total merged 
numbers) with $t_{\rm sf}$. 
Equivalently, the figures show us the age distribution 
of present new born, alive and merged DDs. 
Since the age of a DD is defined as $t_{\rm disc}-t_{\rm sf}$ (\S\ref{sec_ps}), 
the age distribution is obtained by transforming the x-scale to $t_{\rm disc}-t_{\rm sf}$.

The top panels in Fig.\,\ref{fig_cddot} indicates that 
the present new-born He+He and CO+He DDs come mainly from early star
formation $t_{\rm sf} < 8 $ Gyr. This is not the case for CO+CO and
ONeMg+X DDs. 
The top panels in Fig.\,\ref{fig_cddot} also show that the present new-born CO+CO and ONeMg+X come almost entirely
from recent star formation; $8 < t_{\rm sf} < 10$ Gyr. 

This remarkable discrepancy is basically a result of $t_{\rm MS}$. 
Figure\,\ref{fig_progenitor} (right panel) illustrates that a massive MS+MS binary generally 
takes less than a few Myr and 1 Gyr to form ONeMg+X and CO+CO DDs respectively, 
leading to their formation much more quickly than He+He and CO+He DDs. This also 
results in the birth rates of ONeMg+X and CO+CO increasing dramatically 
at an early stage of disc evolution. Only after some time do 
CO+He and He+He DDs emerge and do their birth rates grow to reach the present values shown in the left 
panel of Fig.\,\ref{fig_birnum}. Due to the 
quasi-exponential SF rate, the birth rates of ONeMg+X and CO+CO DDs start to decline after 
reaching a peak value. The long $t_{\rm MS}$ for single He white dwarfs also results in a small 
contribution of star formation from 0$-$7.5 Gyr to the number of
new-born ONeMg+He DDs at the present. 
During the same period, there are no new-born CO+CO DDs. 

Figure\,\ref{fig_cdot} shows that the {\it contribution} of 
star formation to the present numbers (top panels) and total merged 
numbers (bottom panels) of all types of DDs decreases
monotonically as a function of $t_{\rm sf}$. 
This is because most DDs from all epochs survive to the
present time due to their wide orbital separations. 
The number of He+He DDs from each epoch decreases with $t_{\rm sf}$
more sharply than for CO+CO DDs, although early star formation 
provides more He+He DDs than CO+CO DDs. 
A similar situation arises for CO+He and ONeMg+He DDs. 
This result is consistent with stellar evolution and the assumed SF rate.

The significance of computing the present number of DDs using
Eq.\,\ref{eq_number1}, which represents the sum of DDs arising from
different star-formation epochs, is that it demonstrates the link
between the SF history of the galaxy (or, at least, the thin disc in the 
present investigation) and 
the distribution of the properties of present-day DDs, which can be
deduced from, for example, their gravitational wave signal. 

Figure\,\ref{fig_birnum} shows the variation of $\nu$, $\zeta$ and $n_{\rm dd}$ 
(Eq.\,\ref{eq_number1}) of 
different types of DD with age $t_{\rm disc}$. 
The individual properties of the current $n_{\rm dd}$ DDs 
will be used to calculate the gravitational wave signal \citep{Yu11}. 

\subsection{The structure of the thin disc and the local density of DDs}
\label{sec_tn}

The use of a realistic disc model is important in order to describe the
distance distribution of white dwarf binary systems from the Sun. 
\citet{Sackett97} proposed a double exponential distribution. 
\citet{Phleps00} derived three functions for the star density 
distribution in their model of a thin disc plus thick disc 
(exponential $+$ exponential, hyperbolic secant $+$ exponential, 
and squared hyperbolic secant $+$ exponential, respectively) from 
fits to deep star counts carried out in the Calar Alto Deep Imaging 
Survey.  

We here model the thin disc in the Galaxy using a squared hyperbolic 
secant plus exponential distribution expressed as:
\begin{equation}
\rho_{\rm d}(R,z) = \frac{M_{\rm tn}}{4\pi h_{R}^{2}h_{z}} e^{-R/h_{R}}
\textrm{sech}^{2}(-z/h_{z})  \quad {\rm M_{\odot} pc^{-3}},
\end{equation}
where $R$ and $z$ are the natural cylindrical coordinates of the
axisymmetric disc, $h_{R}=2.5$\,kpc is the scale length of the
disc, and $h_{z}=0.352$\,kpc is the scale height of the thin disc. 
$M_{\rm tn}$ is the mass of the thin disc, 
which is determined by the star formation rate. We adopt the position of the Sun 
to be $R_{\rm sun} = 8.5$\,kpc, $z_{\rm sun}$ = $16.5$\,pc \citep{Freudenreich98}. 
We neglect the age and mass dependence of the scale height. This thin-disc model 
is consistent with the model of \citet{Klypin02} and \citet{Robin03}, and also 
in agreement with Hipparcos results and the observed rotation curve.

From the SF rate, the total mass of stars in the thin disc at age 10 Gyr 
is $M_{\rm tn}\approx$ 5.2$\times10^{10}~M_{\odot}$. Combining the thin disc 
model and the mass of stars in the thin disc, the stellar density in the solar 
neighbourhood is $6.27\times10^{-2}{\rm M_{\odot}pc^{-3}}$ for the thin disc, 
These values are consistent with the Hipparcos result, ($7.6\pm1.5)\times10^{-2}~
{\rm M_{\odot}pc^{-3}}$ \citep{Creze98} and the dynamical structure of 
the thin disc \citep{Klypin02,Robin03}. The local density of DDs in the model is 
$1.98\times10^{-4}$ $\rm pc^{-3}$.

\section{Impact of star formation on the DD population}
\label{sec_sfimpact}

\begin{figure*}
\centering
\includegraphics[width=18cm,clip]{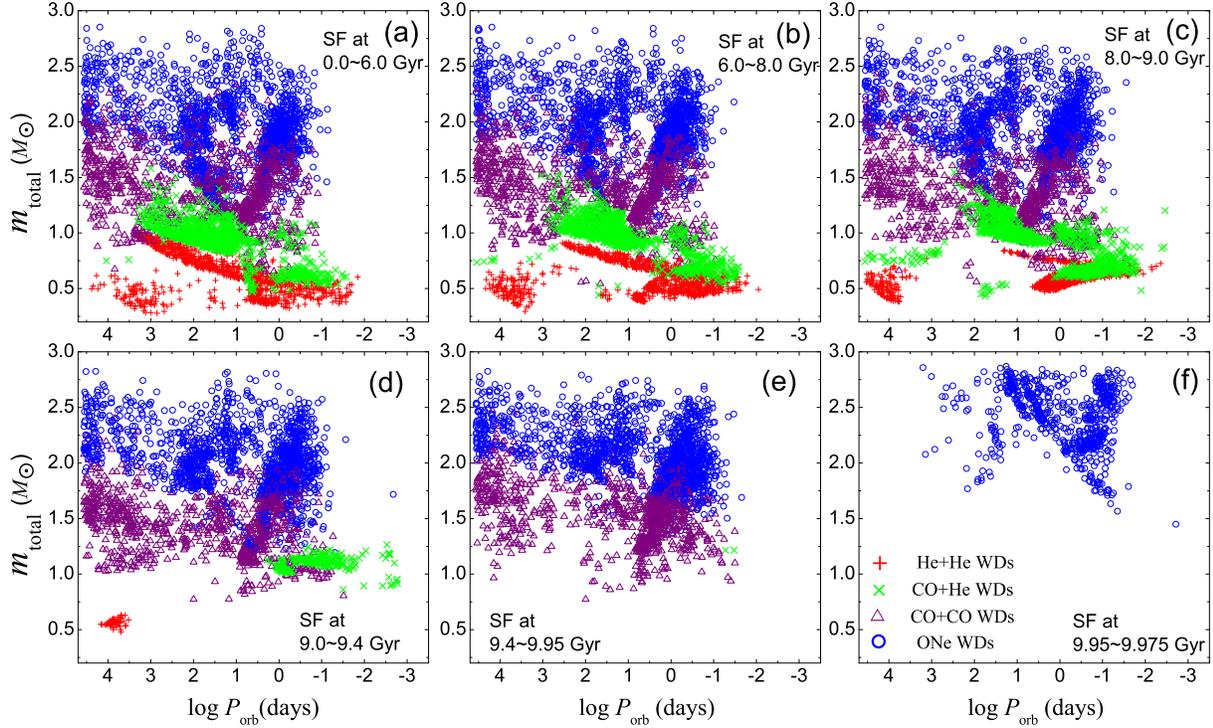}
\caption{The influence of different stages of star formation on the orbital periods 
and the total mass (primary mass + secondary mass) of present DDs.}
\label{fig_mp}
\end{figure*}

The presence of unevolved DDs in Fig.\,\ref{fig_progenitor} indicates the existence of 
a critical time when the first DD of each type was just born. 
For ONeMg+X, CO+CO, CO+He, and He+He DDs respectively, the times are 25 Myr, 
50 Myr, 560 Myr, and 650 Myr. 
Figure\,\ref{fig_mp} illustrates 
the contribution of different epochs of star formation to the
present-day distribution of total mass and orbital periods of the DDs. 
We here distinguish the star formation for the current 
thin disc in six stages, which are 
(a) $0 \leqslant t_{\rm sf}/\rm Gyr < 6$ , 
(b) $6 \leqslant t_{\rm sf}/\rm Gyr < 8$ , 
(c) $8 \leqslant t_{\rm sf}/\rm Gyr < 9$, 
(d) $9 \leqslant t_{\rm sf}/\rm Gyr < 9.4$, 
(e) $9.4 \leqslant t_{\rm sf}/\rm Gyr < 9.95$, and
(f) $9.95 \leqslant t_{\rm sf}/\rm Gyr < 9.75$,  
with the current disc age assumed to be 10 Gyr. 
We do not obtain any DD for star formation taking place after 9.975 Gyr. 
Figure\,\ref{fig_mp} shows that the formation of very close
compact binaries ($\log f>-2.5$) is sensitive to star formation 
between 8 and 9.95 Gyr after the thin disc formed, 
which means that these DDs are most likely to be young. 
Their MS+MS progenitors formed between 50 Myr and 2000 Myr ago.

The total stellar mass formed during the time represented by each panel of 
Fig.\,\ref{fig_mp} is 
$3.62$, $0.85$, $0.38$, $0.15$, $0.19$, and $0.0087\times10^{10}$ $M_{\odot}$ (a to f). 
The current number of DDs in the thin disc derived from each star formation epoch 
in the figure is given in table \ref{tab_DDsf}. These numbers indicate that, 
for current He+He and CO+He DDs, a large number (93.8\% and 87.8\%) have 
ages greater than 4 Gyr, while only 71.9\% and 67.8\% of CO+CO and ONeMg+X have 
ages in the same range. A significant number (3.1\% and 6.7\%) of 
CO+CO and ONeMg+X DDs have been produced by the last 1 Gyr of star formation. 
The number of He+He and CO+He DDs from this period is negligible. 
However, \citet{Yu11} show that DDs with ages less than 2 Gyr 
would contribute substantially to the amplitude of the gravitational
wave signal in several frequency bands. 

\begin{table}
%\begin{minipage}[t]{\columnwidth}
\caption{The contributions of different star formation stages to the numbers of DDs at present day in the quasi-exponential star formation 
model in the thin disc.}
\label{tab_DDsf}
\begin{center}
\begin{tabular}{lccccccc}
\hline
                      & He+He     & CO+He      &  CO+CO     &  ONeMg+X   \\
 \hline
 0$-$6 Gyr         &  48559102 &  19070461  &  51364818  &  5818348      \\
 6$-$8 Gyr         &  3027804  &  2401504   &  12752445  &  1522955    \\
 8$-$9 Gyr         &  178166   &  233402    &  5365912   &  695988     \\
 9$-$9.4 Gyr       &  458      &  1630      &  1347366   &  256836     \\
 9.4$-$9.95 Gyr    &  0        &  5         &  774012    &  309854    \\
 9.95$-$9.975 Gyr  &  0        & 0          &  0         &  309     \\
 9.975$-$10 Gyr    &  0        & 0          &  0         &  0     \\
\hline
 0$-$10 Gyr        &  51765530  & 21707002   & 71604553  &  8604290  \\
\hline
\end{tabular}
\end{center}
%\end{minipage}
\end{table}

According to the classification of the star formation stages and the critical time for the 
birth of DDs, we can see from Fig.\,\ref{fig_mp} that the MS+MS progenitors of CO+CO DDs 
are formed before $t_{\rm sf}\approx9.95$ Gyr. The youngest CO+He DD has an age of about 
560 Myr, but the majority of their MS progenitors formed $>$600 Myr ago. These results are 
consistent with stellar evolution calculations.

Note that in the stellar evolution model a fraction of ONeMg white dwarfs become neutron stars 
and stellar-mass black holes due 
to accretion-induced collapse. 
These do not form type Ia supernovae and are not otherwise considered in our results. 

\section{Comparison of different star formation models}
\label{sec_comparison}

\begin{figure}
\centering
\includegraphics[width=10.5cm,clip,bb=40 20 610 320,angle=0]{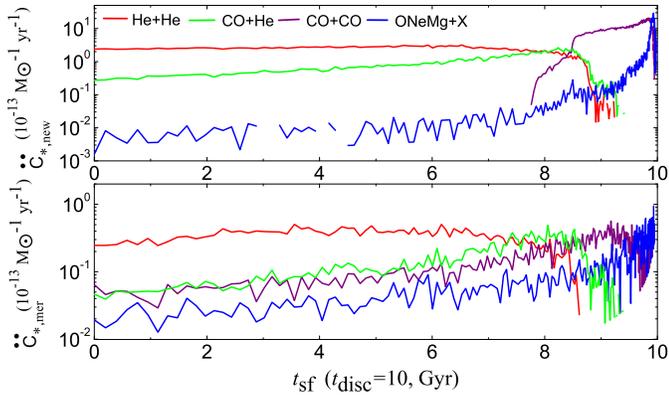}
\caption{The rate contribution function of star formation epochs 
to the present birth rates 
($\ddot{C}_{\ast,\rm new}$, top) and 
merger rates ($\ddot{C}_{\ast,\rm mer}$, bottom) of DDs with 
respect to star formation time $t_{\rm sf}$ in sample.}
\label{fig_cddots}
\end{figure}
\begin{figure}
\centering
\includegraphics[width=9.5cm,clip,angle=0]{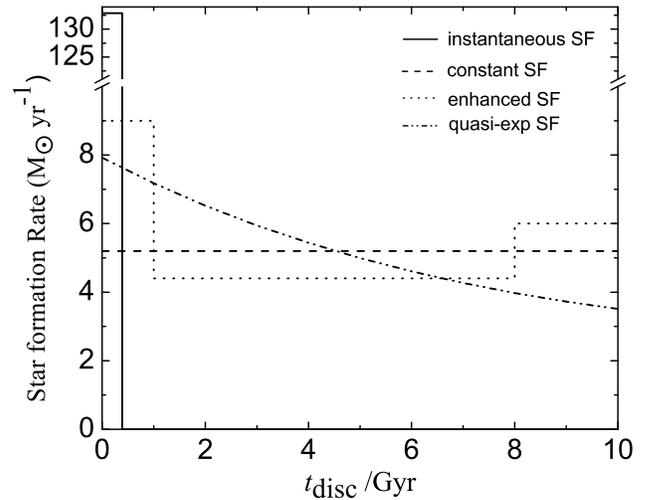}
\caption{Star formation rates in star formation models.}
\label{fig_sfr}
\end{figure}
\begin{figure}
\centering
\includegraphics[width=11cm,clip,bb=25 20 530 230,angle=0]{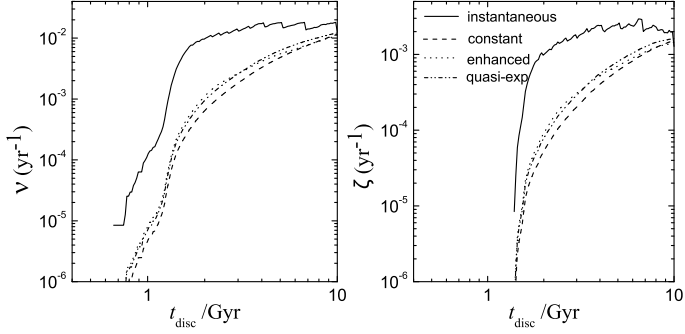}
\caption{Birth rates ($\nu$, left) and merger rates ($\zeta$, right) of He+He DDs in the different star formation models.}
\label{fig_hehesf}
\end{figure}
\begin{figure}
\centering
\includegraphics[width=11cm,clip,bb=25 20 530 230,angle=0]{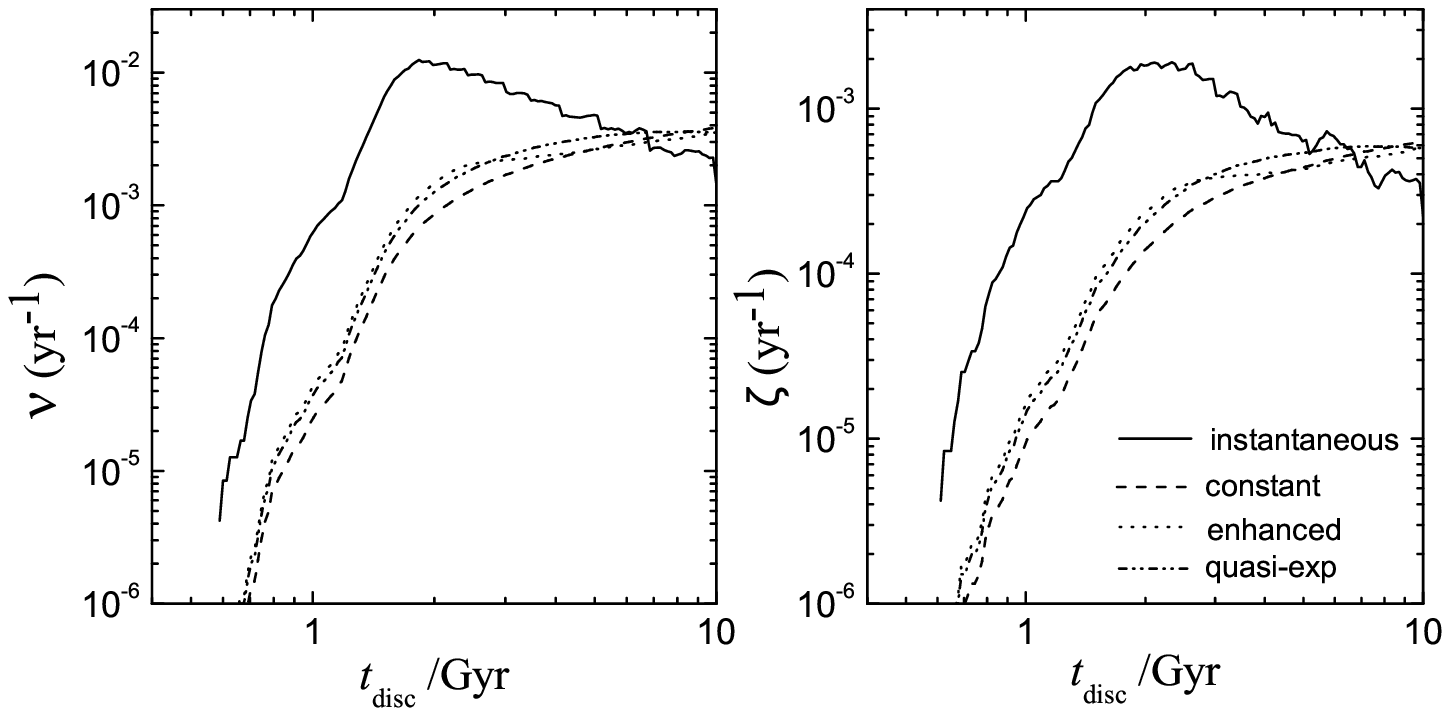}
\caption{As Fig.\,\ref{fig_hehesf} for CO+He DDs.}
\label{fig_cohesf}
\end{figure}
\begin{figure}
\centering
\includegraphics[width=11cm,clip,bb=25 20 530 230,angle=0]{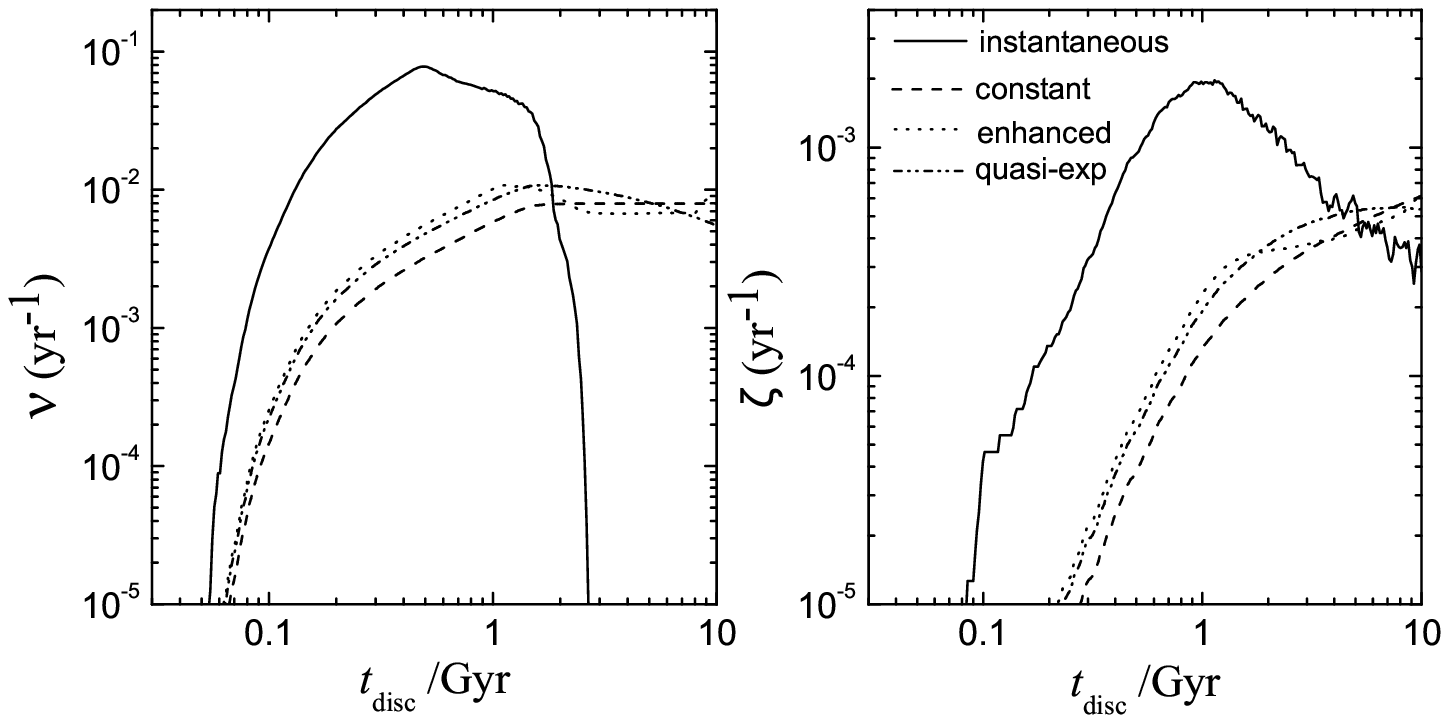}
\caption{As Fig.\,\ref{fig_hehesf} for CO+CO DDs.}
\label{fig_cocosf}
\end{figure}
\begin{figure}
\centering
\includegraphics[width=11cm,clip,bb=25 20 530 240,angle=0]{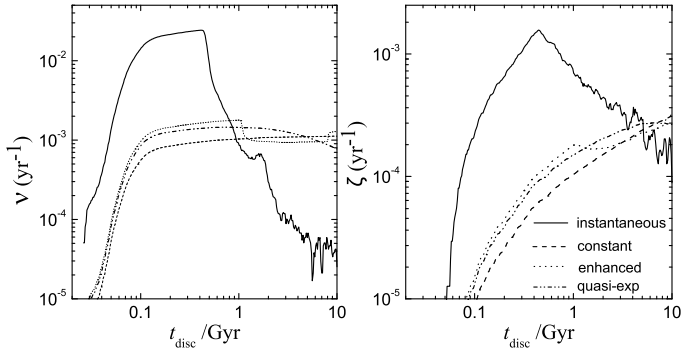}
\caption{As Fig.\,\ref{fig_hehesf} for ONeMg+X DDs.}
\label{fig_othersf}
\end{figure}

We have simulated the present DD population with a quasi-exponential declining SF rate. 
In order to see the influence on the DDs from different SF models, we have arbitrarily constructed 
another three simple SF models and calculated the birth rates, merger rates, and the number of DDs. 
In addition to the quasi-exponential SF model described above, these are :\\
$~~~${\it Instantaneous SF}. A single star burst takes place only at the formation of the thin disc with
\begin{equation}
S=\left\{ 
\begin{array}{c}
M_{\rm tn}/t_{0},~~0\leqslant t_{\rm disc}\leqslant t_{0},\\
~~~~~~~~~~~~~~~~~~~~0,~~t_{0}<t_{\rm disc}\leqslant 10~ {\rm Gyr}.
\end{array}
\right. 
\label{eq_model1}
\end{equation} 
We take $t_{0}= 391$ Myr, so the resulting SF rate in this model is 132.9 
$M_{\odot}$ yr$^{-1}$ on average from $t_{\rm sf}=0$ to 391 Myr. There is no subsequent star formation. \\
$~~~${\it Constant SF}. A constant SF rate at an average rate of  
\begin{equation}
S=5.2~~M_{\odot} ~{\rm yr}^{-1},~~~~0\leqslant t_{\rm disc}\leqslant 10~ {\rm Gyr}.
\label{eq_model2}
\end{equation} \\
$~~~${\it Enhanced SF}. An enhanced SF rate during $t_{\rm disc}$ from 8 to 10 Gyr, 
\begin{equation}
S=\left\{ 
\begin{array}{c}
9,~~0\leqslant t_{\rm disc}\leqslant 1~ {\rm Gyr},\\
4.4,~~1< t_{\rm disc}\leqslant 8~ {\rm Gyr},\\
6,~~8< t_{\rm disc}\leqslant 10~ {\rm Gyr}.
\end{array}
\right. 
\label{eq_model3}
\end{equation} \\

Each model produces a present stellar mass of $5.2\times10^{10}$ $M_{\odot}$ in the thin disc with present age 
$t_{\rm disc}=10$ Gyr in agreement with the dynamic mass of the thin disc \citep{Klypin02}.

Figures\,\ref{fig_cddots} and \ref{fig_sfr} show the star-formation contribution function $\ddot{C}_{\ast}$ 
in our sample to the birth rates and merger rates of the DDs and the SF rate for each SF model. These two 
variables are needed to calculate the birth rates and merger rates in the thin disc in Eq.\,\ref{eq_birthrate}. 
The many small oscillations on the star-formation contribution functions in Fig.\,\ref{fig_cddots} 
(and Fig.\,\ref{fig_cddot}) are due to the statistical noise of the Monte Carlo simulations.

Figures \ref{fig_hehesf} to \ref{fig_othersf} show the birth rates ($\nu$) and merger rates ($\zeta$) of the 
four types of DD in the four models. We see that the instantaneous SF differs greatly from the other 
three models. As a result of the impact of a high SF rate at the formation of the thin disc, $\nu$ and $\zeta$ 
in the instantaneous SF model increase very sharply to reach a maximum value several times greater than in all 
other models. $\nu$ and $\zeta$ subsequently decrease slowly for the He+He DDs 
since they have a long $t_{\rm MS}$. In the cases of the other three types of DD in the instantaneous SF model, 
the decline of $\nu$ and $\zeta$ is faster than He+He DDs due to the shorter $t_{\rm MS}$. When $t_{\rm disc}>2.3$ Gyr, 
roughly the longest $t_{\rm MS}$ of CO+CO DD, there are no more new-born CO+CO DDs. ONeMg+X DDs show a reduction in 
$\nu$ and $\zeta$, except for the ONeMg+He DDs which make a residual contribution to $\nu$ as the disc continue to evolve. 
Again, some oscillations are seen in the birth and merger 
rates in the instantaneous SF model. These are due to statistical
noise from the contribution functions produced by the Monte Carlo
simulations, and numerical noise from the quadrature of Eq.\,\ref{eq_birthrate}.

A big difference for $\nu$ and $\zeta$ in the other three models can also be seen from the figures. Enhanced SF 
causes the present $\nu$ and $\zeta$ of CO+CO and ONeMg+X DDs to be slightly higher than constant SF, while the quasi-exponential 
SF can produce a higher $\nu$ and $\zeta$ of CO+CO and ONeMg+X DDs than enhanced SF when $t_{\rm disc}<4$ Gyr. When 
$t_{\rm disc}>4$ Gyr, we find $\nu$ and $\zeta$ of CO+CO and ONeMg+X DDs decrease dramatically in the quasi-exponential SF model, 
while the values of $\nu$ and $\zeta$ in the constant SF model remain almost constant and higher than the $\nu$ and $\zeta$ 
in the quasi-exponential SF model, up to $t_{\rm disc}=10$ Gyr. The 
significant decreases and increases of the values of $\nu$ and $\zeta$ of CO+CO and ONeMg+X DDs in the enhanced SF model 
are the response of these variables to sudden changes in the SF rate. The present values of $\nu$ and $\zeta$ are 
similar because we adopt a similar value of the average SF rate at the present age of the thin disc in these three models.

\begin{figure}
\centering
\includegraphics[width=11cm,clip,bb=25 20 530 230,angle=0]{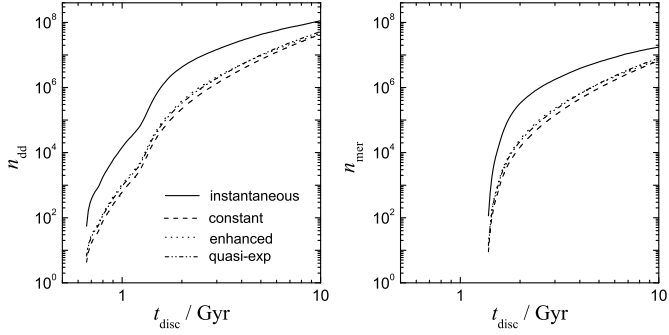}
\caption{Present number ($n_{\rm dd}$, left) and total merger number 
($n_{\rm mer}$, right) of 
He+He DDs in the different star formation models.}
\label{fig_hehenum}
\end{figure}

\begin{figure}
\centering
\includegraphics[width=11cm,clip,bb=25 20 530 230,angle=0]{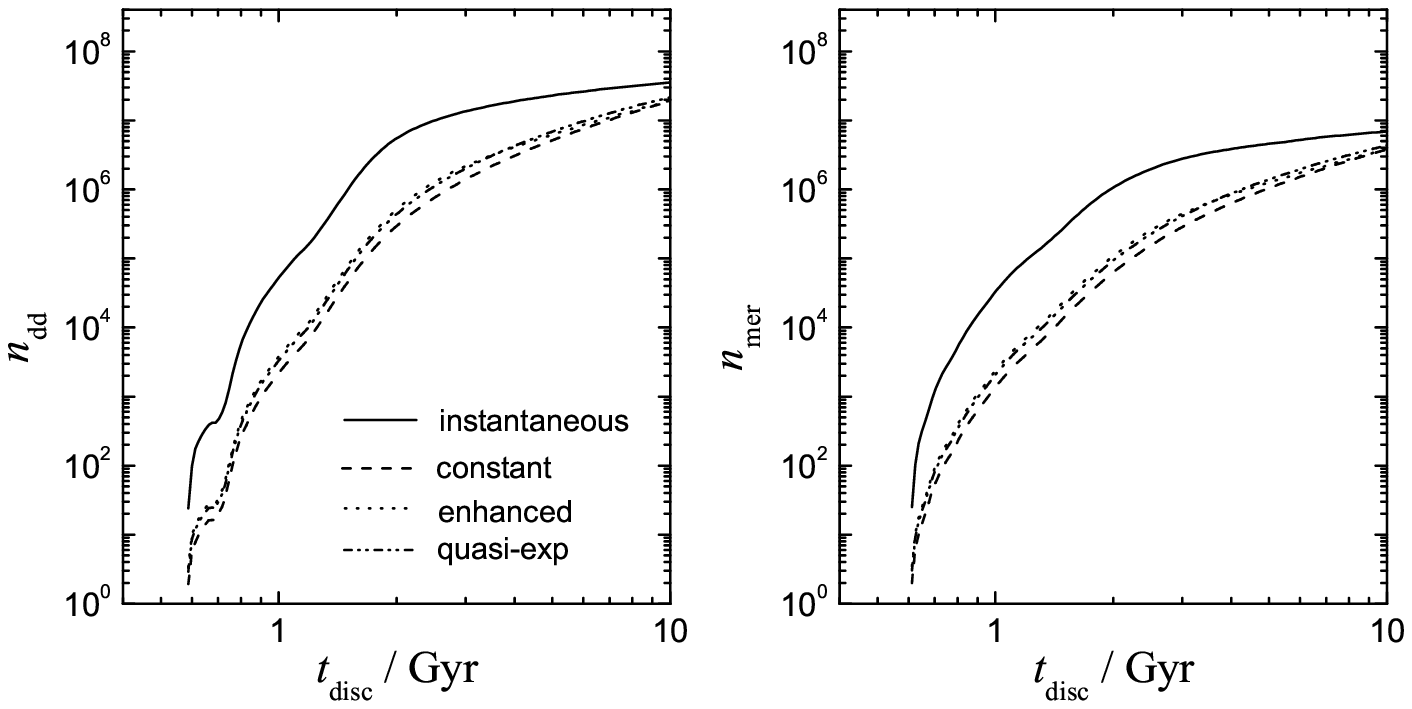}
\caption{As Fig.\,\ref{fig_hehenum} for CO+He DDs.}
\label{fig_cohenum}
\end{figure}

\begin{figure}
\centering
\includegraphics[width=11cm,clip,bb=25 20 530 230,angle=0]{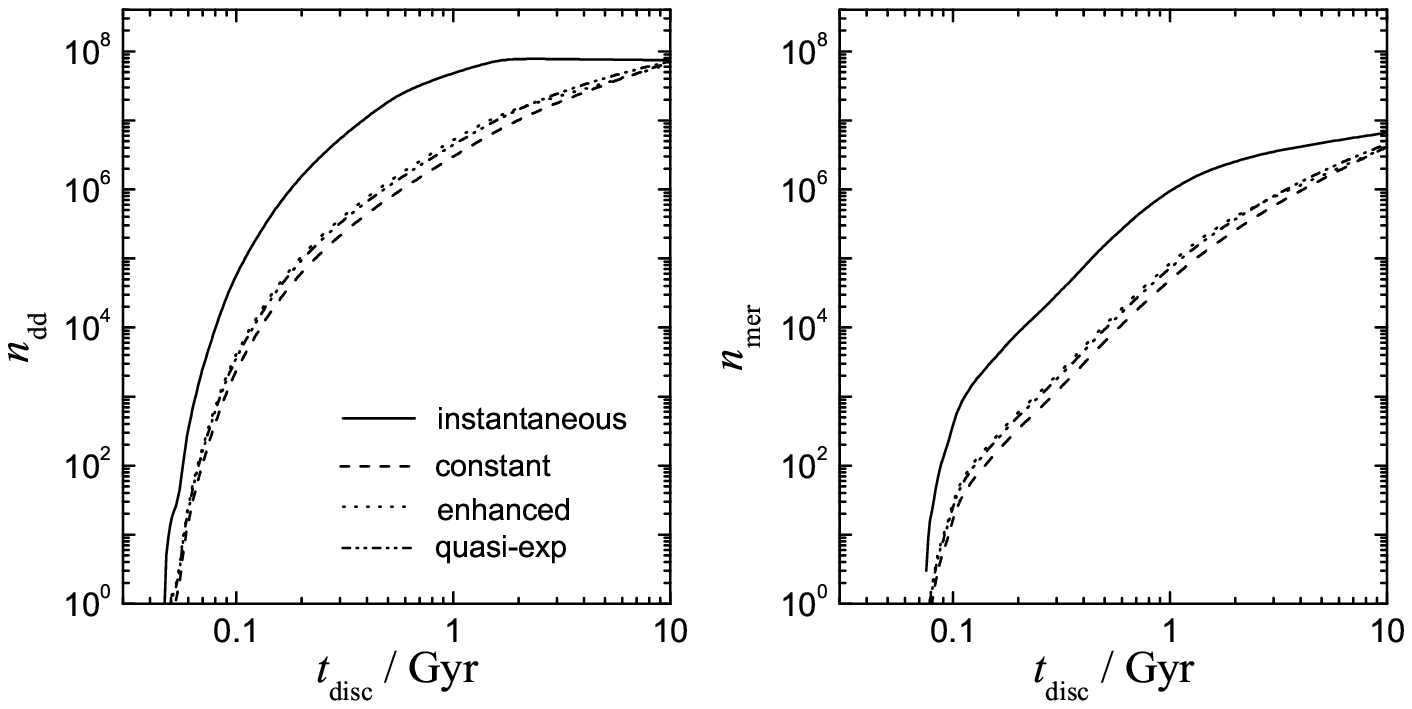}
\caption{As Fig.\,\ref{fig_hehenum} for CO+CO DDs.}
\label{fig_coconum}
\end{figure}

\begin{figure}
\centering
\includegraphics[width=11cm,clip,bb=25 20 530 240,angle=0]{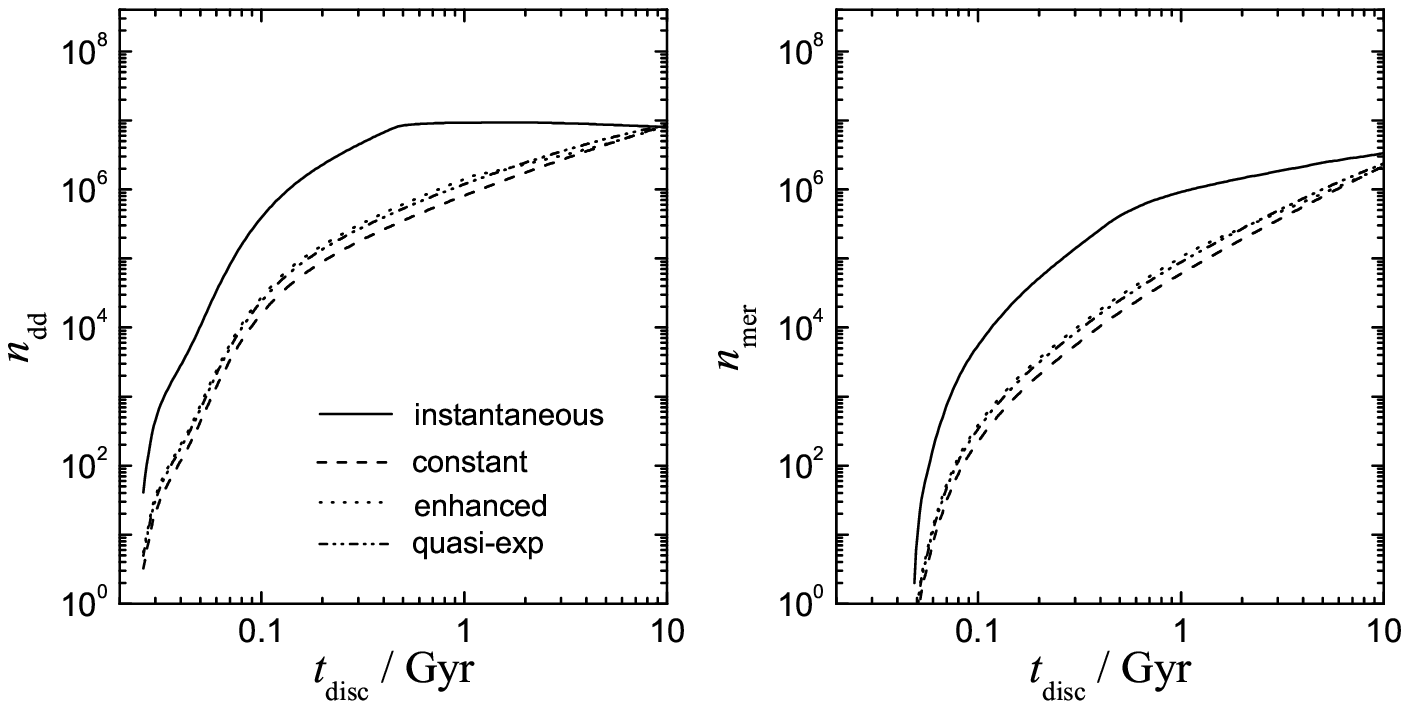}
\caption{As Fig.\,\ref{fig_hehenum} for ONeMg+X DDs.}
\label{fig_othernum}
\end{figure}

The present number and the total merger number of different types of DDs are shown in Figs. \ref{fig_hehenum} 
to \ref{fig_othernum}. A similar evolutionary history of the numbers for various types of DD is generated by the continuous 
SF models since the average values of the SF rate are similar in these SF models. The instantaneous 
SF model produces a slightly higher present numbers of DDs than the other three SF models, and has a distinctly 
different evolution. 

A delay time, which represents the time from when the first DD was born to when the first DD merged, can 
certainly be found in each model if we compare the left panel and the right panel in each figure. The delay time can 
not be less than the shortest life time $t_{\rm DD}$ of a DD, and is affected significantly by stellar evolution 
models and population synthesis parameters ({\it e.g.} the IMF). Since we use the same stellar evolution model and 
population synthesis parameters for each single epoch of the SF in our simulations, the delay time is a constant 
for each SF model. We list the delay time for different types of DD in Table \ref{tab_delayt}. 


\begin{table}
%\begin{minipage}[t]{\columnwidth}
\caption{As table\,\ref{tab_DDsf} except for the instantaneous star formation model.}
\label{tab_DDsfm1}
\begin{center}
\begin{tabular}{lccccccc}
\hline
                 & He+He     & CO+He      &  CO+CO     &  ONeMg+X   \\
 0$-$0.391 Gyr   & 112790783  & 35296981   & 74199554  &  7979607  \\
 0.391$-$10 Gyr  & 0         & 0          & 0         &  0        \\
\hline
 0$-$10 Gyr   &  112790783  & 35296981   & 74199554  &  7979607  \\
\hline
\end{tabular}
\end{center}
%\end{minipage}
\end{table}

\begin{table}
%\begin{minipage}[t]{\columnwidth}
\caption{As table\,\ref{tab_DDsf} except for the constant star formation model. }
\label{tab_DDsfm2}
\begin{center}
\begin{tabular}{lccccccc}
\hline
                 & He+He     & CO+He      &  CO+CO     &  ONeMg+X   \\
 \hline
 0$-$6 Gyr         &  39518987 &  15920357  &  43845056  &  4979173      \\
 6$-$8 Gyr         &  2484043  &  2889597   &  15507241  &  1852669    \\
 8$-$9 Gyr         &  1370423  &  312505    &  7286013   &  946315     \\
 9$-$9.4 Gyr       &  649      &  2309      &  1918541   &  366020     \\
 9.4$-$9.95 Gyr    &  0        &  8         &  1129931   &  454258    \\
 9.95$-$9.975 Gyr  &  0        & 0          &  1         &  462     \\
 9.975$-$10 Gyr    &  0        & 0          &  0         &  0     \\
\hline
 0$-$10 Gyr        &  43374102  & 19124776   & 69686783  &  8598897  \\
\hline
\end{tabular}
\end{center}
%\end{minipage}
\end{table}

\begin{table}
%\begin{minipage}[t]{\columnwidth}
\caption{As table\,\ref{tab_DDsf} except for the enhanced star formation model.  }
\label{tab_DDsfm3}
\begin{center}
\begin{tabular}{lccccccc}
\hline
                 & He+He     & CO+He      &  CO+CO     &  ONeMg+X   \\
 \hline
 0$-$6 Gyr         &  42952616 &  16651902  &  44547319  &  5041578      \\
 6$-$8 Gyr         &  3094596  &  2472828   &  13270620  &  1585457    \\
 8$-$9 Gyr         &  267637   &  351727    &  8315871   &  1080840     \\
 9$-$9.4 Gyr       &  748      &  2666      &  2213701   &  422330     \\
 9.4$-$9.95 Gyr    &  0        &  7         &  1303767   &  524144    \\
 9.95$-$9.975 Gyr  &  0        & 0          &  1         &  533     \\
 9.975$-$10 Gyr    &  0        & 0          &  0         &  0     \\
\hline
 0$-$10 Gyr        &  46315597  & 19479130   & 69651279  &  8654882  \\
\hline
\end{tabular}
\end{center}
%\end{minipage}
\end{table}

\begin{table}
%\begin{minipage}[t]{\columnwidth}
\caption{The delay time, which is from the time the first DD was born to the time the first merged DD appears, for each type of DD.}
\label{tab_delayt}
\begin{center}
\begin{tabular}{lccccccc}
\hline
                    &  He+He    &  CO+He     &  CO+CO     &  ONeMg+X   \\
 \hline
 delay time (Myr)   &  724.03 &  23.94  &  28.65  &  22.39      \\
\hline
\end{tabular}
\end{center}
%\end{minipage}
\end{table}

\section{Supernovae}
\label{sec_snia}

\begin{figure}
\centering
\includegraphics[width=10cm,clip,bb=25 20 530 240,angle=0]{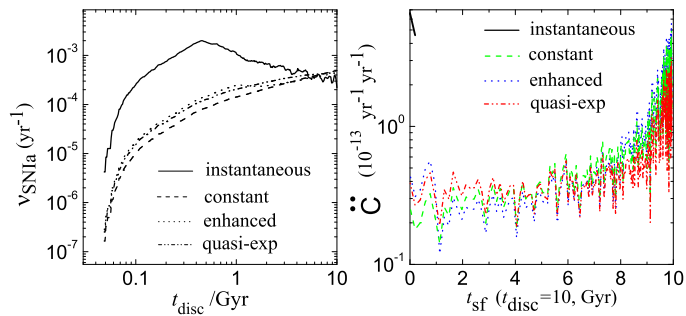}
\caption{The rates of type Ia supernovae from double degenerate mergers channel 
in different star formation models (left panel), and the rate contribution function 
to the prensent-day ($t_{\rm disc}=10$Gyr) supernovae rates in the thin disc 
(right panel).}
\label{fig_snia}
\end{figure}

We have been concerned about the details of birth rate and number of DDs in the present simulation. 
However, we suggest that Eq.\,\ref{eq_birthrate} (to calculate the birth rate) and Eq.\,\ref{eq_number1} 
and \ref{eq_number2} (to calculate the number) are valid for almost any type of star. The only 
thing required is to determine the corresponding contribution of each SF epoch to the numbers of the 
stars in question with respect to the variation of the SF ({\it i.e.} $t_{\rm sf}$). In terms of the determination 
of the contribution, it is important to determine all possible formation channels. Otherwise we 
should specify that the birth rate or number correspond to a specific formation channel. We would like to 
take two important examples, type Ia supernovae (SNIa) and core collapse supernovae (ccSN, including type Ib/Ic and 
type II). 

Previous studies ({\it e.g.} \citet{Iben84,Tutukov92,Iben97a,Yungelson98,Yungelson00,Dahlen04}) indicate that 
mergers of DDs with a total mass exceeding the Chandrasekhar mass limit ($M_{\rm ch}$) would be one formation 
channel to produce SNIa. We here investigate the birth rate of SNIa from the mergers of DDs, and we 
constrain the progenitors to be those DDs with total mass $>M_{\rm ch}=1.38~M_{\odot}$ 
which can merge in the time $t_{\rm disc}$ to $t_{\rm disc}+\delta t_{\rm disc}$. 

The birth rates of SNIa for the four SF models are shown in the left panel in Fig.\,\ref{fig_snia}, and the 
rate contribution function from each single epoch in the four models is shown in the right panel in the same 
figure. The calculation of the rate contribution function is from the 
number of DDs satisfying the conditions to become SNIa in our simulation. Note that the curve in the right panel 
is also the age distribution of SNIa, taking into account the SF rate in the thin disc. This figure indicates that 
both old and young DDs contribute to the present SNIa rates. 

From the left panel, we see that the birth 
rate of SNIa from the instantaneous SF model reaches its maximum value very quickly with the evolution of the thin disc. 
The maximum value is significantly higher than that in another three SF models since the initial SF rate in the instantaneous SF 
model is more than $\sim$30 times that in other SF models. The enhanced SF model produces the highest present rate of 
SNIa in the four SF models. This is because the present SF contributes the majority of mergers. Again, from Fig.\,\ref{fig_snia}, 
continuous SF is also the reason why the rate of SNIa keeps growing with the the evolution of the thin disc. The present-day SNIa rate 
is about 2.19, 4.92, 5.07, and 4.34 $\times10^{-4}$ yr$^{-1}$ for the instantaneous, constant, enhanced, and quasi-exponential SF 
models respectively.

We have to emphasize that we here only take into account the merger channel for the rate of thin-disc SNIa. This 
leads to our results being less than the occurrence rate inferred from observations, $\sim4\times10^{-3}$ yr$^{-1}$ 
\citep{Cappellaro97}, as a few other formation channels would contribute to the observed occurrence rate \citep{Yungelson00,Han04,Hachisu08}. 

We note that the occurrence (or birth) rate of SNIa at $t_{\rm
    disc}=10$ Gyr in our simulations is smaller than in some other
recent theoretical studies ({\it e.g.} \citet{Ruiter09,Mennekens10}) by a factor of $1.5-4$. Since all these studies 
(including ours) adopt the same IMF for the primary main sequence
stars, the same flat distribution for initial mass ratio, and the same 
initial eccentricity distribution, we infer that the differences
between our study and others are mainly caused by the initial 
distribution of orbital separations and by the treatment of mass transfer and mass loss during binary evolution. 

Both \citet{Ruiter09} and \citet{Mennekens10} adopt a distribution of the orbital separation 
$\frac{{\rm d} n}{{\rm d} a}\propto a^{-1}$ \citep{abt83} 
when $a\leqslant10^{5}$ $R_{\odot}$, while we use a distribution of $\propto a^{0.2}$ \citep{Griffin85,Han98} for $a\leqslant10$ $R_{\odot}$, 
and $\propto a^{-1}$ for $10\leqslant a\leqslant5.75\times10^{6}$ $R_{\odot}$. This may lead to the generation 
of more close DDs in their models than our model.

For common-envelope ejection, we adopt the $\gamma$-algorithm
  \citep{Nelemans05} and fix $\gamma=1.5$. Other studies use the
  $\alpha$-algorithm \citep{Webbink84}.
Increasing the common-envelope ejection efficiency (for example,
inreasing $\alpha\lambda=0.5\rightarrow1.0$) results in more binaries 
going through a double common-envelope phase to become close DDs, and
hence forming potential SNIa progenitors. A close comparison 
between the studies implies that the common-envelope ejection
efficiency is lower in our calculation than in others. In the most
extreme case, the \citet{Yungelson10} SNIa birth rate from the DD
merger channel is higher than ours by a factor
of $\sim$10 because they adopted a very high 
common-envelope ejection efficiency ($\alpha\lambda=2.0$) and a 
higher SF rate ($\sim$8 $M_{\odot}$yr$^{-1}$). 

A significant difference between the studies concerns the 
  star-formation history for early-type (elliptical)
  galaxies.  
  Both \citet{Ruiter09} and \citet{Mennekens10} approximate 
  ``instantaneous star formation'' by a  $\delta$-function
  star-burst. 
  Our approximation of ``instantaneous SF'' assumes that the
  initial star-burst lasted a few million years at a constant rate. 
  This leads to the peak value of the SNIa rate occurring rather 
  later in our models than in the studies in \citet{Ruiter09} and
  \citet{Mennekens10}. However, none of these approximations may
  reflect the true SF history of early-type galaxies; recent studies
  indicate that bright early-type galaxies show signs of 
  current star formation \citep{Yi05} and, by inferrence, may have
  experienced intermittent star-burst episodes. 

The case of ccSN is easier than that of SNIa. Their progenitors, main sequence stars with mass approximately greater than 
8 $M_{\odot}$, evolve very quickly on a timescale $\lesssim$20 Myr. This is even shorter than the $t_{\rm MS}$ 
of ONeMg+X (see Fig.\,\ref{fig_progenitor}), so all the new-born ccSN results from SF after a few tens Myr (see criterion in \S\ref{sec_ps}). 
Then the SF rate can be regarded as a constant $S(t_{\rm disc}$). From Eq.\,\ref{eq_birthrate}, we have the birth rate of the supernovae 
$SNR(t_{\rm disc})\approx[N_{\rm ccSN}/m_{\rm total}]\cdot S(t_{\rm disc})$, where $N_{\rm ccSN}$ is the number of stars going 
supernova in a calculation of a sample of main sequence stars, $m_{\rm total}$ represents the total mass of the sample 
, and $t_{\rm disc}=10$ Gyr. Using the same parameters as in the simulation in this paper, we obtain 
$N_{\rm ccSN}/m_{\rm total}\approx4.5\times10^{-3}$ $M_{\odot}^{-1}$ and so $SNR(10)\approx$ 2.34, 2.7 and 1.58 
century$^{-1}$ for the three continuous SF models. These values are consistent with the observations of 1.2 $-$ 3 
century$^{-1}$ \citep{Smith78,vandenbergh94,McKee97,Timmes97,Diehl06} in terms of the mass of $^{26}$Al, the number 
of massive stars in the localized HII region and the extragalactic supernova scaled calculation.

\section{Discussion}
\label{sec_summary}

Our simulation is based only on theory with some simple assumptions. The final 
outputs, the rates and numbers, are derived from two key inputs which are the SF rate 
and the contribution functions for the evolved stars. On the other hand, if we know the 
rates and numbers of objects in a galaxy, especially some types of exotic stars, 
we are able to do a deconvolution by setting the observed rate as an input parameter 
to derive the SF rate at specific epochs. In fact, this method is used to obtain the current 
SF rate via the observation of supernovae \citep{Heger03,Diehl06} and a global SF rate by 
observations of late type main sequence stars \citep{Rocha-Pinto00}. The contribution 
functions provide an important link between SF history and the diverse stellar components 
in different galaxies, since they reflect the sensitivity of a group of the same type of stars 
to the SF history of a galaxy. 

The time grid $\delta t_{\rm sf}$ (or $\delta t_{\rm disc}$) can cause some uncertainties. For the 
quasi-exponential declining SF rate, another computation with a low-resolution time grid of $\delta t_{\rm sf}=0.026$ 
results in positive deviations from 
the high resolution time grid ($\delta t_{\rm sf}=0.00867$) of 1.03\%, 0.82\%, 0.04\% and 0.21\% for the birth rates 
of He+He, CO+He, CO+CO and ONeMg+X DDs respectively; the deviations of the numbers are 3.97\%, 2.53\%, 1.13\% and 0.87\% 
for the different types of DDs. In the new grid, the deviations of the numbers calculated 
from Eq.\,\ref{eq_number1} compared with Eq.\,\ref{eq_number2} are 4.7\%, 2.8\%, 1.5\% and 1.4\% for each type of DD. 
The time error for different stellar evolution phase in our simulation is less than 4.2\%.

\section{Conclusion}
\label{sec_conclusion}

In this paper 
we have investigated the relation between the evolution of birth rates, merger rates and numbers of different types 
of DD and the SF history in the thin disc of the Milky Way Galaxy. By analysing how a quasi-exponentially 
declining SF rate influences the rates and observed numbers of DDs with respect to the evolution of the disc and SF, 
we find that SF between 0 and $\sim$8 Gyr dominates the present rates and total numbers of He+He DDs and CO+He DDs. 
Similarly, the current numbers of CO+CO and ONeMg+X DDs mainly come from early SF, although the later SF ({\it e.g.} 4 to 8 Gy) 
contributes more CO+CO than it does for He+He which are the two largest DD population in the thin disc. 

However, the present birth and merger rates of CO+CO and ONeMg+X DDs are strongly governed by the recent star formation 
({\it i.e.} 8 to 9.95 Gyr). SF before $\approx$7.5 Gyr does {\it not} contribute to the present birth rates of CO+CO ang ONeMg+X 
DDs, but it has some contribution to the merger rates. More importantly, with the SF history related 
population synthesis approach in this paper, not only are we able to determine the rates and numbers of DDs, 
but we can also obtain the distributions of properties of current DD population from different stages of SF. 

We have compared the impact of different SF models, namely the instantaneous, the constant, the enhanced, and the 
quasi-exponential SF, on the rates and numbers of DDs and the rates of two types of supernovae. The evolution 
tracks of the rates and numbers from the four models are quite different. A distinct difference can be found between the instantaneous 
model and the other three continuous SF models. This model gives historical rates and numbers obviously higher than the other three models. 
However, the present rates of DDs from this model are apparently lower than the other three models. 

In addition to the DDs, we have calculated the rates of SNIa and ccSN. The evolution of the rates of SNIa is basically similar in the four SF models, 
but the instantaneous model can produce a higher rate in the past because of the very high SF rate at the formation of the thin disc. The present rates of 
SNIa are 2.19, 4.92, 5.07, and 4.34 $\times10^{-4}$ yr$^{-1}$ for the instantaneous, the constant, the enhanced, and the quasi-exponential SF models respectively. 
The rates of ccSN from all four SF models, $\approx$1.5 to 3 century$^{-1}$, are consistent with the observations. 

\section*{Acknowledgments}
The Armagh Observatory is supported by a grant from the Northern Ireland 
Dept. of Culture Arts and Leisure. The authors are grateful to Prof. Lev Yungelson 
for useful suggestions, and to colleagues at the Armagh Observatory for their stimulating 
discussion. We thank the referee for the useful suggestions and comments. 

\bibliographystyle{mn2e}
\bibliography{ddsfh}

\begin{thebibliography}{}

\bibitem[\protect\citeauthoryear{{Abt}}{{Abt}}{1983}]{abt83}
{Abt} H.~A.,  1983, ARA\&A, 21, 343

\bibitem[\protect\citeauthoryear{{Cappellaro}, {Turatto}, {Tsvetkov},
  {Bartunov}, {Pollas}, {Evans} \& {Hamuy}}{{Cappellaro}
  et~al.}{1997}]{Cappellaro97}
{Cappellaro} E.,  {Turatto} M.,  {Tsvetkov} D.~Y.,  {Bartunov} O.~S.,  {Pollas}
  C.,  {Evans} R.,    {Hamuy} M.,  1997, A\&A, 322, 431

\bibitem[\protect\citeauthoryear{{Creze}, {Chereul}, {Bienayme} \&
  {Pichon}}{{Creze} et~al.}{1998}]{Creze98}
{Creze} M.,  {Chereul} E.,  {Bienayme} O.,    {Pichon} C.,  1998, A\&A, 329,
  920

\bibitem[\protect\citeauthoryear{{Cropper}, {Harrop-Allin}, {Mason}, {Mittaz},
  {Potter} \& {Ramsay}}{{Cropper} et~al.}{1998}]{Cropper98}
{Cropper} M.,  {Harrop-Allin} M.~K.,  {Mason} K.~O.,  {Mittaz} J.~P.~D.,
  {Potter} S.~B.,    {Ramsay} G.,  1998, MNRAS, 293, L57

\bibitem[\protect\citeauthoryear{{Dahlen}, {Strolger}, {Riess}, {Mobasher},
  {Chary}, {Conselice}, {Ferguson}, {Fruchter}, {Giavalisco}, {Livio}, {Madau},
  {Panagia} \& {Tonry}}{{Dahlen} et~al.}{2004}]{Dahlen04}
{Dahlen} T.,  {Strolger} L.,  {Riess} A.~G.,  {Mobasher} B.,  {Chary} R.,
  {Conselice} C.~J.,  {Ferguson} H.~C.,  {Fruchter} A.~S.,  {Giavalisco} M.,
  {Livio} M.,  {Madau} P.,  {Panagia} N.,    {Tonry} J.~L.,  2004, ApJ, 613,
  189

\bibitem[\protect\citeauthoryear{{Diehl}, {Halloin}, {Kretschmer}, {Lichti},
  {Sch{\"o}nfelder}, {Strong}, {von Kienlin}, {Wang}, {Jean}, {Kn{\"o}dlseder},
  {Roques}, {Weidenspointner}, {Schanne}, {Hartmann}, {Winkler} \&
  {Wunderer}}{{Diehl} et~al.}{2006}]{Diehl06}
{Diehl} R.,  {Halloin} H.,  {Kretschmer} K.,  {Lichti} G.~G.,
  {Sch{\"o}nfelder} V.,  {Strong} A.~W.,  {von Kienlin} A.,  {Wang} W.,  {Jean}
  P.,  {Kn{\"o}dlseder} J.,  {Roques} J.,  {Weidenspointner} G.,  {Schanne} S.,
   {Hartmann} D.~H.,  {Winkler} C.,    {Wunderer} C.,  2006, Nature, 439, 45

\bibitem[\protect\citeauthoryear{{Evans}, {Iben} \& {Smarr}}{{Evans}
  et~al.}{1987}]{Evans87}
{Evans} C.~R.,  {Iben} I.~J.,    {Smarr} L.,  1987, ApJ, 323, 129

\bibitem[\protect\citeauthoryear{{Farmer} \& {Roelofs}}{{Farmer} \&
  {Roelofs}}{2010}]{Farmer10}
{Farmer} A.,  {Roelofs} G.,  2010, ArXiv e-prints

\bibitem[\protect\citeauthoryear{{Freudenreich}}{{Freudenreich}}{1998}]{Freude%
nreich98}
{Freudenreich} H.~T.,  1998, ApJ, 492, 495

\bibitem[\protect\citeauthoryear{{Gokhale}, {Peng} \& {Frank}}{{Gokhale}
  et~al.}{2007}]{Gokhale07}
{Gokhale} V.,  {Peng} X.~M.,    {Frank} J.,  2007, ApJ, 655, 1010

\bibitem[\protect\citeauthoryear{{Griffin}}{{Griffin}}{1985}]{Griffin85}
{Griffin} R.~F.,  1985, in {P.~P.~Eggleton \& J.~E.~Pringle} ed., NATO ASIC
  Proc. 150: Interacting Binaries {The distributions of periods and amplitudes
  of late-type spectroscopic binaries}.
pp 1--12

\bibitem[\protect\citeauthoryear{{Hachisu}, {Kato} \& {Nomoto}}{{Hachisu}
  et~al.}{2008}]{Hachisu08}
{Hachisu} I.,  {Kato} M.,    {Nomoto} K.,  2008, ApJ, 683, L127

\bibitem[\protect\citeauthoryear{{Han}}{{Han}}{1998}]{Han98}
{Han} Z.,  1998, MNRAS, 296, 1019

\bibitem[\protect\citeauthoryear{{Han} \& {Podsiadlowski}}{{Han} \&
  {Podsiadlowski}}{2004}]{Han04}
{Han} Z.,  {Podsiadlowski} P.,  2004, MNRAS, 350, 1301

\bibitem[\protect\citeauthoryear{{Heger}, {Fryer}, {Woosley}, {Langer} \&
  {Hartmann}}{{Heger} et~al.}{2003}]{Heger03}
{Heger} A.,  {Fryer} C.~L.,  {Woosley} S.~E.,  {Langer} N.,    {Hartmann}
  D.~H.,  2003, ApJ, 591, 288

\bibitem[\protect\citeauthoryear{{Hurley}, {Pols} \& {Tout}}{{Hurley}
  et~al.}{2000}]{Hurley00}
{Hurley} J.~R.,  {Pols} O.~R.,    {Tout} C.~A.,  2000, MNRAS, 315, 543

\bibitem[\protect\citeauthoryear{{Hurley}, {Tout} \& {Pols}}{{Hurley}
  et~al.}{2002}]{Hurley02}
{Hurley} J.~R.,  {Tout} C.~A.,    {Pols} O.~R.,  2002, MNRAS, 329, 897

\bibitem[\protect\citeauthoryear{{Iben} Jr. \& {Tutukov}}{{Iben} \&
  {Tutukov}}{1984}]{Iben84}
{Iben} Jr. I.,  {Tutukov} A.~V.,  1984, ApJ, 54, 335

\bibitem[\protect\citeauthoryear{{Iben} Jr. \& {Tutukov}}{{Iben} \&
  {Tutukov}}{1997}]{Iben97a}
{Iben} Jr. I.,  {Tutukov} A.~V.,  1997, ApJ, 491, 303

\bibitem[\protect\citeauthoryear{{Klypin}, {Zhao} \& {Somerville}}{{Klypin}
  et~al.}{2002}]{Klypin02}
{Klypin} A.,  {Zhao} H.,    {Somerville} R.~S.,  2002, ApJ, 573, 597

\bibitem[\protect\citeauthoryear{{Kroupa}, {Tout} \& {Gilmore}}{{Kroupa}
  et~al.}{1993}]{Kroupa93}
{Kroupa} P.,  {Tout} C.~A.,    {Gilmore} G.,  1993, MNRAS, 262, 545

\bibitem[\protect\citeauthoryear{{Majewski}}{{Majewski}}{1993}]{Majewski93}
{Majewski} S.~R.,  1993, ARA\&A, 31, 575

\bibitem[\protect\citeauthoryear{{Marsh}, {Nelemans} \& {Steeghs}}{{Marsh}
  et~al.}{2004}]{Marsh04}
{Marsh} T.~R.,  {Nelemans} G.,    {Steeghs} D.,  2004, MNRAS, 350, 113

\bibitem[\protect\citeauthoryear{{McKee}}{{McKee}}{1989}]{McKee89}
{McKee} C.~F.,  1989, ApJ, 345, 782

\bibitem[\protect\citeauthoryear{{McKee} \& {Williams}}{{McKee} \&
  {Williams}}{1997}]{McKee97}
{McKee} C.~F.,  {Williams} J.~P.,  1997, ApJ, 476, 144

\bibitem[\protect\citeauthoryear{{Mennekens}, {Vanbeveren}, {De Greve} \& {De
  Donder}}{{Mennekens} et~al.}{2010}]{Mennekens10}
{Mennekens} N.,  {Vanbeveren} D.,  {De Greve} J.~P.,    {De Donder} E.,  2010,
  A\&A, 515, A89+

\bibitem[\protect\citeauthoryear{{Nelemans} \& {Tout}}{{Nelemans} \&
  {Tout}}{2005}]{Nelemans05}
{Nelemans} G.,  {Tout} C.~A.,  2005, MNRAS, 356, 753

\bibitem[\protect\citeauthoryear{{Nelemans}, {Yungelson} \& {Portegies
  Zwart}}{{Nelemans} et~al.}{2001}]{Nelemans01b}
{Nelemans} G.,  {Yungelson} L.~R.,    {Portegies Zwart} S.~F.,  2001, A\&A,
  375, 890

\bibitem[\protect\citeauthoryear{{Nelemans}, {Yungelson} \& {Portegies
  Zwart}}{{Nelemans} et~al.}{2004}]{Nelemans04}
{Nelemans} G.,  {Yungelson} L.~R.,    {Portegies Zwart} S.~F.,  2004, MNRAS,
  349, 181

\bibitem[\protect\citeauthoryear{{Phleps}, {Meisenheimer}, {Fuchs} \&
  {Wolf}}{{Phleps} et~al.}{2000}]{Phleps00}
{Phleps} S.,  {Meisenheimer} K.,  {Fuchs} B.,    {Wolf} C.,  2000, A\&A, 356,
  108

\bibitem[\protect\citeauthoryear{{Popper}}{{Popper}}{1980}]{Popper80}
{Popper} D.~M.,  1980, ARA\&A, 18, 115

\bibitem[\protect\citeauthoryear{{Ramsay}, {Hakala}, {Wu}, {Cropper}, {Mason},
  {C{\'o}rdova} \& {Priedhorsky}}{{Ramsay} et~al.}{2005}]{Ramsay05}
{Ramsay} G.,  {Hakala} P.,  {Wu} K.,  {Cropper} M.,  {Mason} K.~O.,
  {C{\'o}rdova} F.~A.,    {Priedhorsky} W.,  2005, MNRAS, 357, 49

\bibitem[\protect\citeauthoryear{{Robin}, {Reyl{\'e}}, {Derri{\`e}re} \&
  {Picaud}}{{Robin} et~al.}{2003}]{Robin03}
{Robin} A.~C.,  {Reyl{\'e}} C.,  {Derri{\`e}re} S.,    {Picaud} S.,  2003,
  A\&A, 409, 523

\bibitem[\protect\citeauthoryear{{Rocha-Pinto}, {Scalo}, {Maciel} \&
  {Flynn}}{{Rocha-Pinto} et~al.}{2000}]{Rocha-Pinto00}
{Rocha-Pinto} H.~J.,  {Scalo} J.,  {Maciel} W.~J.,    {Flynn} C.,  2000, A\&A,
  358, 869

\bibitem[\protect\citeauthoryear{{Ruiter}, {Belczynski}, {Benacquista},
  {Larson} \& {Williams}}{{Ruiter} et~al.}{2010}]{Ruiter10}
{Ruiter} A.~J.,  {Belczynski} K.,  {Benacquista} M.,  {Larson} S.~L.,
  {Williams} G.,  2010, ApJ, 717, 1006

\bibitem[\protect\citeauthoryear{{Ruiter}, {Belczynski} \& {Fryer}}{{Ruiter}
  et~al.}{2009}]{Ruiter09}
{Ruiter} A.~J.,  {Belczynski} K.,    {Fryer} C.,  2009, ApJ, 699, 2026

\bibitem[\protect\citeauthoryear{{Sackett}}{{Sackett}}{1997}]{Sackett97}
{Sackett} P.~D.,  1997, ApJ, 483, 103

\bibitem[\protect\citeauthoryear{{Smith}, {Biermann} \& {Mezger}}{{Smith}
  et~al.}{1978}]{Smith78}
{Smith} L.~F.,  {Biermann} P.,    {Mezger} P.~G.,  1978, A\&A, 66, 65

\bibitem[\protect\citeauthoryear{{Timmes}, {Diehl} \& {Hartmann}}{{Timmes}
  et~al.}{1997}]{Timmes97}
{Timmes} F.~X.,  {Diehl} R.,    {Hartmann} D.~H.,  1997, ApJ, 479, 760

\bibitem[\protect\citeauthoryear{{Tutukov}, {Yungelson} \& {Iben}
  Jr.}{{Tutukov} et~al.}{1992}]{Tutukov92}
{Tutukov} A.~V.,  {Yungelson} L.~R.,    {Iben} Jr. I.,  1992, ApJ, 386, 197

\bibitem[\protect\citeauthoryear{{van den Bergh} \& {McClure}}{{van den Bergh}
  \& {McClure}}{1994}]{vandenbergh94}
{van den Bergh} S.,  {McClure} R.~D.,  1994, ApJ, 425, 205

\bibitem[\protect\citeauthoryear{{Webbink}}{{Webbink}}{1984}]{Webbink84}
{Webbink} R.~F.,  1984, ApJ, 277, 355

\bibitem[\protect\citeauthoryear{{Wielen}, {Jahrei{\ss}} \&
  {Kr{\"u}ger}}{{Wielen} et~al.}{1983}]{Wielen83}
{Wielen} R.,  {Jahrei{\ss}} H.,    {Kr{\"u}ger} R.,  1983, in {A.~G.~D.~Philip
  \& A.~R.~Upgren} ed., IAU Colloq. 76: Nearby Stars and the Stellar Luminosity
  Function {The Determination of the Luminosity Function of Nearby Stars}.
pp 163--170

\bibitem[\protect\citeauthoryear{{Yi}, {Yoon} \& {Kaviraj}}{{Yi}
  et~al.}{2005}]{Yi05}
{Yi} S.~K.,  {Yoon} S.-J.,    {Kaviraj} S. e.~a.,  2005, ApJ, 619, L111

\bibitem[\protect\citeauthoryear{{Yu} \& {Jeffery}}{{Yu} \&
  {Jeffery}}{2010}]{Yu10}
{Yu} S.,  {Jeffery} C.~S.,  2010, A\&A, 521, A85+

\bibitem[\protect\citeauthoryear{{Yu} \& {Jeffery}}{{Yu} \&
  {Jeffery}}{2011}]{Yu11}
{Yu} S.,  {Jeffery} C.~S.,  2011, Submitted

\bibitem[\protect\citeauthoryear{{Yungelson} \& {Livio}}{{Yungelson} \&
  {Livio}}{1998}]{Yungelson98}
{Yungelson} L.,  {Livio} M.,  1998, ApJ, 497, 168

\bibitem[\protect\citeauthoryear{{Yungelson}}{{Yungelson}}{2010}]{Yungelson10}
{Yungelson} L.~R.,  2010, Astronomy Letters, 36, 780

\bibitem[\protect\citeauthoryear{{Yungelson} \& {Livio}}{{Yungelson} \&
  {Livio}}{2000}]{Yungelson00}
{Yungelson} L.~R.,  {Livio} M.,  2000, ApJ, 528, 108

\end{thebibliography}

\label{lastpage}
\end{document}